\documentclass[12pt,preprint]{aastex}
\usepackage{psfig}
\newcommand{\instinit}{
 \setcounter{footnote}{1}
}
\newcommand{\instnew}[1]{
  \value{footnote}    
  \addtocounter{footnote}{-1}  
  \refstepcounter{footnote}
  \label{#1}
 }
\newcommand{\instend}{
  \setcounter{footnote}{0}
}
\begin{document}


\shorttitle{\textit{HST}/STIS observations of \objectname[]{GRB~000301C}}
\shortauthors{Smette et al.}
\title{\textit{HST}/STIS observations of \objectname[]{GRB~000301C}: \linebreak
CCD imaging and NUV MAMA  spectroscopy$^*$
}
\author{
Alain Smette\altaffilmark{\ref{gsfc},\ref{noao},\ref{fnrs}},
Andrew S. Fruchter\altaffilmark{\ref{stsci}},
Theodore R. Gull\altaffilmark{\ref{gsfc}},
Kailash C. Sahu\altaffilmark{\ref{stsci}},
Larry Petro\altaffilmark{\ref{stsci}},
Henry Ferguson\altaffilmark{\ref{stsci}},
James Rhoads\altaffilmark{\ref{stsci}},
Don J. Lindler\altaffilmark{\ref{acc}},
Rachel Gibbons\altaffilmark{\ref{stsci}},
David W. Hogg\altaffilmark{\ref{ias},\ref{hubble}},
Chryssa Kouveliotou\altaffilmark{\ref{marshall},\ref{usra}},
Mario Livio\altaffilmark{\ref{stsci}},
Duccio Macchetto\altaffilmark{\ref{stsci},\ref{esa}},
Mark R. Metzger\altaffilmark{\ref{caltech}},
Holger Pedersen\altaffilmark{\ref{copenhagen}},
Elena Pian\altaffilmark{\ref{itsre}},
Stephen E. Thorsett\altaffilmark{\ref{santacruz},\ref{sloan}},
Ralph A.M.J. Wijers\altaffilmark{\ref{suny}},
Johan P. U. Fynbo\altaffilmark{\ref{aarhus},\ref{eso}},
Javier Gorosabel\altaffilmark{\ref{dsri}},
Jens Hjorth\altaffilmark{\ref{copenhagen}},
Brian L. Jensen\altaffilmark{\ref{copenhagen}},
Alan Levine\altaffilmark{\ref{mit}},
Donald A. Smith\altaffilmark{\ref{mit}},
Tom Cline\altaffilmark{\ref{gsfc}},
Kevin Hurley\altaffilmark{\ref{berkeley}},
Jack Trombka\altaffilmark{\ref{gsfc}}
}
\instinit
\altaffiltext{
  \instnew{gsfc}
}
{ 
NASA Goddard Space Flight Center, Greenbelt MD~20771, USA;\newline
asmette@band3.gsfc.nasa.gov,
gull@sea.gsfc.nasa.gov,
cline@lheavx.gsfc.nasa.gov,
u1jit@lepvax.gsfc.nasa.gov 
}

\altaffiltext{
\instnew{noao}
}
{ 
National Optical Astronomy Observatories, P.O. Box
26732, 950 North Cherry Avenue, Tucson AZ~85726-6732, USA.  
}
\altaffiltext{
\instnew{fnrs}
}
{ 
Collaborateur Scientifique, FNRS, Belgium.  
}
\altaffiltext{
\instnew{stsci}
}
{ 
Space Telescope Science Institute, 3700 San Martin Drive, 
Baltimore MD~21218, USA;
fruchter@stsci.edu,
sahu@stsci.edu,
petro@stsci.edu
ferguson@stsci.edu,
rhoads@stsci.edu,
gibbons@stsci.edu,
livio@stsci.edu,
macchetto@stsci.edu, 
}
\altaffiltext{
\instnew{acc}
}
{ 
Advanced Computer Concepts, Inc./Goddard Space
Flight Center, Code 681, Greenbelt MD~20771, USA;
lindler@rockit.gsfc.nasa.gov 
}
\altaffiltext{
\instnew{ias}
}
{ 
Institute for Advanced Study, Princeton NJ~08540,    USA;
hogg@ias.edu 
}
\altaffiltext{
\instnew{hubble}
}
{ 
Hubble Fellow.  
}
\altaffiltext{
\instnew{marshall}
}
{ 
NASA Marshall Space Flight Center, ES-84,    Huntsville AL~35812, USA;
chryssa.kouveliotou@msfc.nasa.gov 
}
\altaffiltext{
\instnew{usra}
}
{ 
Universities Space Research Association.  
}
\altaffiltext{
\instnew{esa}
}
{ 
Affiliated to the Astrophysics Division, Space  Science Department, 
European Space Agency.  
}
\altaffiltext{
\instnew{caltech}
}
{ 
Department of Astronomy, Caltech, MS~105-24,    Pasadena CA~91125, USA;
mrm@grus.caltech.edu 
}
\altaffiltext{
\instnew{copenhagen}
}
{ 
Astronomical Observatory, University of Copenhagen, 
Juliane Maries Vej 30, D-2100, Copenhagen \O, Denmark; 
holger@ursa.astro.ku.dk,
jens@astro.ku.dk,
brian\_j@astro.ku.dk 
}
\altaffiltext{
\instnew{itsre}
}
{ 
Istituto di Tecnologie e Studio delle Radiazioni  Extraterrestri,
C.N.R., Via Gobetti 101, I--40129 Bologna, Italy; 
pian@tesre.bo.cnr.it 
}
\altaffiltext{
\instnew{santacruz}
}
{ 
Department of Astronomy and Astrophysics,  University of California,
Santa Cruz CA~95064, USA; 
thorsett@ucolick.org 
}
\altaffiltext{
\instnew{sloan}
}
{ 
Alfred P. Sloan Research Fellow.  
}
\altaffiltext{
\instnew{suny}
}
{ 
State University of New York, 452 Earth and Space    Sciences
Building, Stony Brook, NY 11794--3800; 
rwijers@ourania.ess.sunysb.edu 
}
\altaffiltext{
\instnew{aarhus}
}
{ 
Institute of Physics and Astronomy, {\AA}arhus  University, DK--8000
{\AA}arhus C., Denmark; 
jfynbo@ifa.au.dk 
}
\altaffiltext{
\instnew{eso}
}
{ 
European Southern Observatory, D--85748, Garching    bei M{\"u}nchen, Germany 
}
\altaffiltext{
\instnew{dsri}
}
{ 
Danish Space Research Institute, Juliane Maries    Vej 30 DK--2100
Copenhagen {\/O}, Denmark; 
jgu@dsri.dk 
}
\altaffiltext{
\instnew{mit}
}
{ 
Massachusetts Institute of Technology, Center for Space Research, 77
Massachusetts Avenue, Cambridge MA~02139, USA; 
aml@space.mit.edu,
dasmith@space.mit.edu 
}
\altaffiltext{
\instnew{berkeley}
}
{ 
University of California, Berkeley, Space    Sciences Laboratory,
Berkeley CA~94720--7450, USA; 
khurley@ssl.berkeley.edu
    \newline $^*$ Based on observations with the NASA/ESA Hubble Space
    Telescope, obtained at the Space Telescope Science Institute,
    which is operated by the Association of Universities for Research
    in Astronomy, Inc. under NASA contract No. NAS5-26555.
\instend
}
\begin{abstract}
  We present \textit{HST}/STIS observations of the optical counterpart
  (OT) of the $\gamma$-ray burster \objectname[]{GRB~000301C} obtained
  on 2000 March 6, five days after the burst. CCD clear aperture
  imaging reveals a $R \simeq 21.50\pm0.15$ source with no apparent
  host galaxy. An 8000 s, $1150 < \lambda/\mbox{\AA} < 3300$ NUV--MAMA
  prism spectrum shows a flat or slightly rising continuum (in
  $f_\lambda$) between 2800 and 3300 \AA, with a mean flux
  $8.7^{+0.8}_{-1.6}\pm2.6~10^{-18}~
  \mathrm{ergs}~\mathrm{s}^{-1}~\mathrm{cm}^{-2}~\mbox{\AA}^{-1}$, and
  a sharp break centered at $2797\pm25$~\AA.  We interpret this as the
  \ion{H}{1} Lyman break at $z = 2.067\pm0.025$ indicating the
  presence of a cloud with a \ion{H}{1} column density
  $\log{(N_\mathrm{HI}~\mathrm{cm}^{2})} > 18$ on the line-of-sight to
  the OT.  This measured redshift is conservatively a lower limit to
  the GRB redshift. However, as all other GRBs which have deep
  \textit{HST} images appear to lie on the stellar field of a host
  galaxy, and as the large \ion{H}{1} column density measured here and
  in later ground-based observations is unlikely on a random
  line-of-sight, we believe we are probably seeing absorption from
  \ion{H}{1} in the host galaxy.  In any case, this represents the
  largest direct redshift determination of a $\gamma$-ray burster to
  date.  Our data are compatible with an OT spectrum represented by a
  power-law with an intrinsic index $\alpha = 1.2$ ($f_{\nu} \propto
  \nu^{-\alpha}$) and no extinction in the host galaxy or with $\alpha
  = 0.5$ and extinction by SMC-like dust in the OT rest-frame with
  $A_\mathrm{V} = 0.15$. The large $N_\mathrm{HI}$ and the lack of a
  detected host are similar to the situation for damped Ly-$\alpha$
  absorbers at $z > 2$.
\end{abstract}

\setcounter{equation}{0}
\section{Introduction}
\label{sec:introduction}

Gamma-ray bursts (hereafter, GRB) remain mysterious nearly 30 years
after their first detection \citep{kle73}, although some progress has
been made in their understanding.
The large sample of events collected by the BATSE experiment on-board
the Compton Gamma-Ray Observatory shows an isotropic angular
distribution \citep{mee92}. However, their intensity
distribution shows fewer weak bursts than expected from a homogeneous
distributions of sources in a Euclidean space. It was therefore thought
that GRBs were  of  extragalactic origin
\citep{mee92}.
 
Rapidly distributed arcminute localizations derived from X-ray
instruments such as those on \textit{BeppoSAX} \citep{cos97} and, more
recently, \textit{RXTE} \citep{smi99} satellites have revealed the
existence of fading afterglows and enabled efficient searches for GRB
counterparts to be carried out at other wavelengths.
The optical transient (hereafter, OT) to GRB~970228 was the first to
be discovered \citep{par97}. Its location angularly close to a faint
diffuse object, possibly a galaxy, suggested that they are associated,
providing further evidence that GRBs lie at cosmological distances
\citep[][later, \citet{fru99a}  showed that the diffuse object
is indeed a $V = 25.8$ galaxy]{par97,slp+97}.  This hypothesis was soon
confirmed when a direct spectrum of the OT to GRB~970508 revealed
absorption lines at the same redshift as the underlying $z = 0.835$
galaxy \citep{met97,bdk+98}.  Since then, OTs have been found for
roughly half of the GRBs for which an X--ray afterglow has been
detected and most of them appear to be associated with faint, $ z>
0.4$ galaxies \citep{hf99,kul00}.  In particular, the probable host
galaxy of GRB~971214 has a redshift of $z = 3.42$ \citep{kul98}.
 
Most popular theories describing the GRB phenomenon
suggest that GRB activity will be
closely tied in time to episodes of massive star formation.
The collapse of massive stars \citep{pac98} requires an active or
recently active star forming host, while most neutron star -- neutron
star or neutron star -- black hole binary mergers \citep{pac91,nar92}
occur not long after star formation.
The evidence for such an environment is already suggestive, as a
number of host galaxies show strong emission lines associated with
star formation \citep{met97,djo98,kul98,blo99,vfr+00}, while direct
imaging with \textit{HST}/STIS and NICMOS indicates that they are
unusually blue with $ V - H < 1$ whereas most galaxies in the Hubble
Deep Field have $V - H > 1$ \citep{fru99b}. We thus expect that the
spectra of OTs will show evidence that the events take place in
environments of massive star formation: large hydrogen column density,
large extinction by dust or even absorption by molecular hydrogen
\citep{dra00}. All lead to features in the rest-frame UV region.

We  have  thus  embarked on  a   program  to   obtain, as targets   of
opportunity, \textit{HST}/STIS  near  ultraviolet spectra of GRB  OTs. 
Although \citet{bsw+97} have suggested using grating modes of STIS for
studying GRBs, we chose to use the  prism as it  is the most efficient
available  spectral  element due   to  its  low dispersion   and broad
spectral  coverage.       Combined  with   the     near   ultraviolet,
photon-counting  multianode microchannel array (hereafter,  NUV--MAMA)
detector,  the prism   mode    provides spectral coverage over     the
wavelength range $1150 \la  \lambda/\mbox{\AA} \la 3300$. However, its
highly non-linear dispersion  makes the wavelength calibration at  the
red end ($2800 \la \lambda/\mbox{\AA} \la 3300$) especially critical.

Here we present \textit{HST}/STIS clear aperture CCD imaging and the
first ultraviolet spectrum of a GRB optical counterpart,
\objectname[]{GRB~000301C}. This burst was detected at
9$^h$51$^m$37$^s$ (UTC) on 2000 March 1 by the ASM instrument aboard
the \textit{RXTE} satellite, as well as by Ulysses and NEAR
\citep{smi00}. Its $\gamma$--ray light curve showed a common, simple
shape: singly peaked, fast--rise followed by a slow decay, lasting a
total of 10 seconds, which qualifies it as a short/intermediate
duration burst \citep{jfg+00}.  The resulting composite localization
of area $\sim50$~arcmin$^2$ was imaged on 2000 March 3.14--3.28 (UTC),
by \citet{fyn00a} with the Nordic Optical Telescope (hereafter, NOT)
which revealed a then $R = 20.09\pm0.04$, blue optical counterpart
\citep{jfg+00}.

All exposures were taken on 2000 March 6 (UTC) during a single, 5
orbit visit. The first orbit was dedicated to imaging, acquisition and
calibration.  Four 2000 s NUV--MAMA prism spectra were obtained during
the subsequent four orbits.

\section{Imaging}
\label{sec:imaging}

The field of GRB 000301C was observed using the STIS 50CCD, clear
aperture mode.  In this setting no filter is interposed between the
CCD and the sky, and the bandpass is determined entirely by the
reflectivity of the optics and the response of the CCD.  Three
exposures of 480 s each were taken in a diagonal dither pattern and
combined using the Drizzle algorithm and associated techniques
\citep{fh99,fru97}.  The drizzled output image, presented in Figure
\ref{fig:image}, has pixels of one-half the linear dimensions of the
original images, or $0 \farcs 025355\pm0.000035$, since the pixel
scale of the CCD is $0 \farcs 05071\pm0.00007$ \citep{mb97}.

At the position of the optical transient of GRB 000301C we find an
unresolved object (FWHM $0 \farcs 087$) whose appearance is consistent
with the STIS CCD Point Spread Function (hereafter, PSF). The mean
count-rate of detected photons within 20 CCD pixels (containing more
than 99.6\% of the flux \citep{l00}) from the OT image centroid is
56.0 counts s$^{-1}$ on mean 2000 March 6.22 (UTC). Assuming a power
law spectrum with $\alpha=-0.8$ and a foreground Galactic extinction
with $A_\mathrm{V} = 0.16$, as determined from the dust extinction map
given by \citet{sfd98}, we obtain a magnitude of $R=21.5\pm0.15$,
where the uncertainty is dominated by  the fact that the STIS CCD
clear aperture mode is not completely calibrated.

Any underlying galaxy would have to have $R \ga 24$, or be unusually
compact for it to avoid detection in our data. We therefore believe
that the apparent flattening -- the decrease of the OT luminosity
fading rate -- 
of the light curve in the $R$--band
reported by many observers before this image was taken (cf.
\citet{gar00,hal00a,ber00,rho01} and references therein) is not due
to an underlying host galaxy. 
Instead, it may either be due to the jet
itself \citep{b+00} or to micro-lensing \citep{gls00}.
The diffuse emission to the northwest of the OT is the galaxy
previously identified as a possible host \citep{rho00}.
From our STIS image, we derived an AB magnitude $25.1\pm0.2$ in the
STIS clear aperture passband. The lack of
emission between the two and the good fit between the OT and the STIS
PSF suggest that the OT host galaxy has not yet been detected.

\section{Spectroscopy}
\label{sec:spectroscopy}

\subsection{Acquisition and reduction}
\label{sec:acq_reduc}

In this section, we describe the acquisition and the reduction
procedures which required special attention. The reduction was
performed using IDL and the GSFC version of CALSTIS \citep{lin99}.

Each TIME--TAG, 2000 s spectrum was obtained with the prism using the
2125 \AA\, setting -- so that the spectral region corresponding to
wavelengths $\sim 2125$ \AA\, falls onto the central region of the
NUV--MAMA detector -- and the $52\arcsec \times 0.5\arcsec$ slit.  After
each scientific spectrum  an automatic wavelength
calibration exposure was taken through the narrower $52\arcsec \times
0.05\arcsec$ slit.

We used the $R = 18.05$ star located $\sim5.7$\arcsec\, west to the OT
(cf.  Fig. \ref{fig:image}) as an offset star.  The offset values were
determined from the $3\times900$ s $R$-band NOT image.  Acquisition
took place after the STIS CCD imagery and just before the NUV--MAMA
prism spectroscopy. This order allowed us to obtain the CCD images
independently of the success of the spectral acquisition. However, as
the acquisition produces a small change in the location of the object
on the detector, an additional offset has to be taken into account for
the zero--point determination of the wavelength calibration (see
below).

Each scientific raw dataset displays a spectrum whose spatial profile
can be modeled by a Gaussian with a mean FWHM of 2.5 pixels, or
0.072\arcsec, as expected from a point source.  There is no
significant variation in the count rate between the different
datasets, as well as no significant variation in the count-rate during
the course of an orbit.

The sum of the sky and dark backgrounds at the location of the OT
spectrum was determined for each individual exposure. It was 
obtained by interpolation based on a fifth degree polynomial fitted
independently to each column of the image, avoiding a 20 pixel wide
region centered on the object spectrum.  Extraction was performed by
summing up 7 rows centered on the peak of the 2-D spectrum.

Each extracted spectrum was then wavelength calibrated. This step is
especially critical, so  we describe it here
in more detail.  

The prism dispersion relation $\lambda = \lambda(X)$ gives the
wavelength $\lambda$ corresponding to the pixel\footnote[1]{Throughout
  this paper, a pixel refers to a native, 1024$\times$1024 format MAMA
  pixel, not a `high-resolution' pixel (cf. \citet{wkb+98,kwb+98}).}
number $X$ in the extracted spectrum. It is well modelled by a
relation
\begin{equation}
\label{eq:prism_dispersion}
\lambda = \sum_{i=0}^4 \frac{a_i}{(X-X_{0})^i}.
\end{equation}
Due to the strong blending of the lines especially at the red end of
the spectrum, no determination of the dispersion solution (i.e., the
values of each coefficient $a_i$) is feasible on-orbit.  Instead, the
values of $a_i$ were obtained during the pre--launch, thermal vacuum
testing of STIS, using an external platinum discharge lamp fed through
a vacuum monochromator, the $52\arcsec \times 0.05\arcsec $ slit and
the same 2125 \AA\, setting as for the observations. The residuals to
the fit were then measured to have an \textit{rms} equal to 0.16
pixels.

The quantity $X_{0}$ is, in pixels, the separation along the
dispersion direction of the projected location of the object on the
detector from the projected location of the center of the $52\arcsec 
\times 0.05\arcsec$  slit used for the determination of the dispersion
solution. $X_0$ determines the zero-point of the wavelength
calibration for each exposure and can be represented as the sum of
three quantities:
\textit{(1)}
$\Delta \phi_\mathrm{row}$,
the difference in pixels between the angular separations along the
dispersion direction between the OT and the offset star as determined
on the STIS CCD and NOT images;
\textit{(2)} 
$\Delta \chi$,
the difference in pixels between the projection on the detector of the
center of the slit used for the wavelength calibration relative to its
location during the pre--launch calibration;
\textit{(3)} 
$\Delta \beta$,
the difference in pixels between the location of the  center of the
slit used during the scientific exposures
($52\arcsec\times0.5\arcsec$) and that used for the automatic
wavelength calibration after each scientific exposure
($52\arcsec\times0.05\arcsec$).
In the following, we describe how each of these three quantities can be
determined.  

The acquisition process usually centers the object in the slit used
for the scientific exposure. However, the nearby $R = 18.05$ bright
star was considered too close to the OT and could have interfered with
the normal procedure.  Instead, it was used as an offset star.
Consequently, the acquisition process (a) changed the spacecraft
orientation so that the offset star  is centered on the slit by
an on-board centroid algorithm, and then (b)
modified the spacecraft pointing to take into account the offsets
in right ascension and declination between the OT and the offset star
as determined on the NOT image. However, the STIS image provides more
accurate, but slightly different values for these offsets, so that we
can determine exactly how far the object actually is from the slit
center.  $\Delta \phi_\mathrm{row}$ represents the difference along
the dispersion direction between the two offset determinations.  Note
that we also checked that the on-board algorithm provided a centroid
location for the offset star on the acquisition image consistent,
within 0.1 CCD pixel (0.2 MAMA pixel), with the one used to determine
the offset between the star and the OT on the STIS CCD image.

Ideally, $\Delta \chi$ is best determined for each scientific dataset
by cross-correlating its associated wavelength calibration exposure
with a pre--launch wavelength calibration exposure.  Unfortunately,
the internal lamps were not operated during the pre-launch tests at
the same current setting as during the on-board calibration, so that
the shape of the lamp spectra were too different to be useful.  In
order to mitigate this effect, we built a template of the prism
wavelength calibration lamp at the same setting as used during our
observations based on G140L, G230L and G430L wavelength calibration
exposures and the relevant sensitivity curves.  However, even this
template does not show exactly the same shape at the red end as the
wavelength calibration spectrum.  The value of $\Delta \chi$ was
finally obtained by cross-correlating a 61-pixel long region (along
the dispersion direction) of each spectrum where the dispersion is
relatively large.  As a check, we measured the center of the
Lyman-$\alpha$ geo-coronal line determined by the middle position
between its edges, and found that it falls within 1/6 of a pixel of
the expected location.  Finally, shifts due to thermal or flexure
causes were measured to be between 0.02 and 0.2 pixels from orbit to
orbit, and thus probably less than 0.1 pixel during an exposure.

Finally, regular on--board STIS calibrations enable us to determine
$\Delta \beta$ to better than 1/20 CCD pixel, (1/10 MAMA pixel).

Consequently, a conservative ($2\sigma$) estimate of the error on the
zero-point of the wavelength calibration is 0.5 MAMA pixels, or 
0.23, 1.4, 5.2, 12, 22 and 29 \AA\, at $\lambda = $ 1200, 1500, 2000,
2500, 3000 and 3300 \AA, respectively.

Since the  dispersion is small at  $\sim 2900$  \AA\, (the location of
the peak), the  effect  of the telescope  and  instrument  line spread
function (hereafter, LSF) must  be  taken  into  account to obtain   a
correct estimate  of   the  count  rate per  pixel.    We,  therefore,
deconvolved  the observed spectrum  using  the Lucy-Richardson method. 
Unfortunately, the blending   of lines in  the  wavelength calibration
spectrum prevents its use to create a LSF.  In the absence of suitable
calibration frames taking into account the effect  of the Point Spread
Function of  the telescope  itself, we use   a LSF based  on the shape
perpendicular to  the  dispersion  of  the 2-D  spectrum   of the flux
standard star \objectname[]{HS 2027+0651}  at  about \mbox{2800 \AA}.  
Ray-tracing studies indicate that this  is a good approximation of the
LSF   along    the   dispersion   direction (C.W.    Bowers,   private
communication).   This  star was observed in  the  prism mode during a
STIS  calibration program.  The  immediate   consequence is that   the
corrected (de-convolved)  mean level  of the count  rate is  larger by
about 25\% between  2800 and 3300  \AA\, compared to the  un-corrected
one.

Before converting  the  count   rate to   flux,  we  have  checked the
wavelength calibration   of  the prism  spectrum of   \objectname[]{HS
  2027+0651} and deconvolved it    before building a  new  sensitivity
curve. Once converted to flux, the 4  individual spectra were rebinned
to a     same set of wavelengths   using   bi-linear interpolation and
averaged.   The final, flux  calibrated spectrum  of \objectname[]{GRB
  000301C} is presented in Figure \ref{fig:simulation}a.

\subsection{Simulations}
\label{sec:simulation}

Since Lucy-Richardson deconvolution of a low S/N spectrum actually
decreases its S/N, we prefer to compare the spectrum in total observed
counts per pixel in wavelength with simulated spectra.  These spectra
were built using an input spectrum expressed in units of flux,
multiplied by the sensitivity curve, integrated over the spectral
range of each pixel, and finally convolved with the LSF. The output
spectrum expressed in count-rate is then multiplied by the exposure
time.

In this paper, we only consider input spectra behaving as power-law of
the form $f_\nu \propto \nu^{-\alpha}$, as expected for GRB afterglows
\citep{mrp94,spn98}, modified to take into account the Galactic
extinction towards \objectname[]{GRB~000301C}, the Lyman-$\alpha$
forest absorption lines, possible absorption due to \ion{H}{1}, H$_2$
as well as extinction by dust located in the host galaxy.

For the Galactic extinction in the direction towards \objectname[]{GRB
  000301C}, we used the value $A_\mathrm{V} = 0.16$ as determined from
the dust extinction map given by \citet{sfd98} and the analytical
expression of the Galactic extinction curve given by \citet{pei92}.
In order to model the Lyman-$\alpha$ forest absorption spectrum, we
first noted that the likely redshift of \objectname[]{GRB~000301C}
(cf.  below) falls in a range ($1.5 \la z \la 2.3$) where the
Lyman-$\alpha$ forest is still poorly studied. So far, the Hubble Deep
Field South (HDF) quasar is the only one in that range whose absorption
spectrum has been observed at resolution high enough to determine the
\ion{H}{1} column density and Doppler parameter by Voigt profile
fitting. We thus used the \ion{H}{1} absorbers listed by \citet{sav99}
to estimate an absorption spectrum due to their respective Lyman
series (Ly-$\alpha$ to Ly-12) lines.  However, we discarded the lines
associated with the  $z=1.942$ Lyman-limit system seen in the HDF
quasar as it is a
single event whose occurrence is dominated by small number statistics.
Naturally, we also eliminated the lines due to clouds whose redshift
is larger than the assumed GRB redshift in the simulation.

Absorption due to \ion{H}{1} located in the host galaxy takes into
account Lyman series lines, modeled by Voigt profiles, as well as
continuous absorption. The redshift of the host galaxy
$z$, the \ion{H}{1} column density $N_\mathrm{HI}$ and
the Doppler parameter $b$ can be easily adjusted.
The modelling of possible absorption due to H$_2$ located in the host
galaxy follows  \citet{dra00}.

Finally, the possible extinction due to dust in the host galaxy,
characterized by the rest-frame $A_\mathrm{V}$ extinction and the type
of dust (Small Magellanic Cloud, Large Magellanic Cloud or
Milky Way) also uses the expression given by  \citet{pei92}.

\subsection{Results}
\label{sec:results_spectroscopy}

Figure \ref{fig:simulation}a reveals a flat or slightly rising
continuum spectrum (in $f_\lambda$) over the range from 2800 \AA\, to
the sensitivity limited red end of the spectrum (3300 \AA).  No useful
constraint on the spectral slope can be derived from the spectrum
alone.  The mean flux is measured to be
$8.7^{+0.8}_{-1.6}\pm2.6~10^{-18}~
\mathrm{ergs}~\mathrm{s}^{-1}~\mathrm{cm}^{-2}~\mbox{\AA}^{-1}$, where
the first error values come from the uncertainties in the wavelength
calibration combined with the strong slope of the sensitivity curve at
$\sim2900$ \AA; the second error value is the standard deviation of
the flux calibrated error array over the relevant pixels.  However, a
sharp break can be seen centered at $2797\pm25$ \AA ($2\sigma$). There
is no significant flux recovery at the blue end of the spectrum.

\subsection{Redshift of GRB~000301C }
\label{sec:redshift}

The observed break is best interpreted as the onset of continuous
absorption below the \ion{H}{1} Lyman break due to the presence of a
cloud with a large neutral hydrogen column density, located at $z =
2.067\pm0.025$ ($2\sigma$) in the line-of-sight to the OT.
From the fact that the observed flux is zero within the error bars
blueward of the break, we determine that $\log
(N_\mathrm{HI} ~\mathrm{cm}^{2}) \ga 18.0$. The decreased sensitivity
at the blue end combined with the possible presence of other
relatively high column density clouds in the Lyman-$\alpha$ forest
does not allow us to set a higher lower limit. An upper limit
$\log{(N_\mathrm{HI} ~ \mathrm{cm}^2)} \le 23.3$ can be set from the
absence of a strong feature at the expected location of the Ly-$\beta$
line at $\sim 3100$ \AA.  

Our redshift determination is confirmed by the presence of weakly or
marginally detected lines, compatible with \ion{Fe}{2}, \ion{Mg}{2}
and other low-ionization species at a redshift $z = 2.0335\pm0.0003$
in a Keck spectrum obtained by \citet{cas00}. More importantly, an
ESO/VLT spectrum covering the wavelength range from 3600 \AA\, to 8220
\AA, was obtained by \citet{jfg+00}. It reveals a damped Ly-$\alpha$
with a large, although uncertain $\log{N_\mathrm{HI}} = 21.2\pm0.5$,
within the range allowed by our spectrum, as well as some associated
\ion{C}{4}, \ion{Fe}{2} and possibly other low-ionization lines. These
lines appear at $z = 2.0404\pm0.0008$, a significantly larger value 
than \citet{cas00} and in reasonable agreement with our value.

The value $z = 2.067\pm0.025$ corresponds to the largest direct
redshift measurement of the OT to a GRB so far. If we assume that the
\ion{H}{1} cloud is associated with the OT host galaxy, then the above
value is the redshift of the OT.  We note however that it is
conservatively a lower limit to the OT redshift.  Indeed, if the
object giving rise to the GRB is located outside of a galactic halo
or, more generally, any \ion{H}{1} cloud, all the absorption lines
detected in the OT spectrum would be due to intervening systems.

Can we set an upper limit to the GRB redshift?  Although the ESO/VLT
spectrum extends down to 3600 \AA, its low signal-to-noise ratio at
the blue end cannot allow us to exclude the presence of low column
density Ly-$\alpha$ forest lines between 3600 \AA\, and $\sim 4000$
\AA. This wavelength corresponds to a firm upper limit of $z = 2.3$ to
the GRB redshift.  This value also corresponds to the maximal redshift
for which our STIS spectrum could reveal Lyman-limit systems which
have $\log{(N_\mathrm{HI} ~ \mathrm{cm}^{2})} > 17$.  Using the number density of
Lyman-limit systems seen in QSO spectra \citep{ste95} in the relevant
redshift range, we estimate that the probability of observing
\textit{no} additional Lyman-limit system over the range $2.067 < z <
2.3$ to be 0.68.  This value is too small to exclude a larger redshift
than $z = 2.067\pm0.025$. Similarly, the probability that a random
line-of-sight covering the redshift range $0 < z < 2.3$ would
encounter at least one damped Ly-$\alpha$ system is $\sim 0.25$ based
on the number density of damped Ly-$\alpha$ sytem determined by
\citet{rt00}. This value is too large to ascertain that the redshift
of the high-column density system corresponds to the redshift of
\objectname[]{GRB~000301C}.  In other words, our \textit{HST}/STIS
spectrum and the ESO/VLT one only allow us to limit with certainty the
redshift range of the OT to be $2.067 < z < 2.3$. However, we can
argue in favor of the $z = 2.067\pm0.025$ value for the redshift of
\objectname[]{GRB~000301C}. Indeed, on one hand, damped Ly-$\alpha$
systems are thought to be mainly caused by merging protogalactic
clumps hosted by collapsed DM halos and, consequently, forming stars
as the progenitors of normal present--day galaxies
\citep{gkh+97,hsr98,hsr00}. On the other hand, most OTs discovered so
far are found to be associated with galaxies \citep{fru99b}, many of
which show signs of significant star formation
\citep{met97,djo98,kul98,fru99a,fru99b,blo99,vfr+00} for which,
presumably, a large reservoir of gas is available. In summary, the
unusually large \ion{H}{1} column density and the fact that most OT
host galaxies found so far appear to actively form stars lead us to
believe that the absorption seen in the spectrum of \objectname[]{GRB
  000301C} is most likely due to the host galaxy.

\subsection{Low extinction towards GRB~000301C}
\label{sec:extinction}

When combined with the contemporaneous  STIS CCD imaging, the STIS spectrum
also allows  us to constrain  the extinction taking place  in the $z =
2.067$  \ion{H}{1}  cloud towards \objectname[]{GRB~ 000301C}.  Figure
\ref{fig:simulation}b compares the observed  total number of counts to
a simulation characterized  by  an intrinsic  power-law  spectrum with
index $\alpha =  0.5$, and  some   extinction in  the host  galaxy  by
SMC-like dust ($A_\mathrm{V}  =   0.15$). However, in  the  wavelength
range of the STIS spectrum, such a model is indistinguishable from one
defined by  an intrinsic power-law  spectrum with index $\alpha = 1.2$
($f_\nu \propto \nu^{-\alpha}$) and no  extinction in the host galaxy. 
These two models not only are able to reproduce  the appearance of our
spectrum without the need of  any molecular hydrogen, but also conform
to the  slope of   the  optical spectrum   \citep{hal00b,fen00,jfg+00}
within its  uncertainty.  They also match the  count-rate  of the STIS
CCD image, within  the 15\%   accuracy  of the IDL  routine  SIM\_STIS
(Plait, private communication), which predicts the expected count-rate
for   the  different  STIS  modes   based   on  the individual optical
components.  However,  if a   steep spectrum with  $\alpha =   1.2$ is
compatible with our STIS imaging  and spectroscopic data alone, a more
shallow one    ($\alpha   \simeq 1$)   is  favored   when ground-based
observations with lower  uncertainties in optical band magnitudes than
the   STIS--derived  $R$  magnitude  are   taken   into  account  (cf. 
\citet{hal00b,rho01}).   Similarly,  $\alpha \la  1$, models with low
extinction $A_\mathrm{V} \la 0.1$   ($A_\mathrm{V} \la 0.05$)   due to
LMC- (Milky  Way-) like dust are  also  consistent with  the STIS data
alone.     They predict  that    the 2175 \AA\,   feature   (cf. e.g.  
\citet{bfo96})  should be  detectable and shifted  to  $\sim6570$ \AA. 
However, its  presence  seems to   be  ruled out by  existing  optical
spectra and photometry \citep{rho01,jfg+00}. Therefore a model with a
relatively small amount ($A_\mathrm{V} \la 0.1$) of SMC-like 
dust appears to best fit the full UV/optical data.

\section{Discussion and conclusion}
\label{sec:conclusion}

The \textit{HST}/STIS  CCD  observations of \objectname[]{GRB~000301C}
revealed a $R =  21.50\pm0.15$ source  on   2000 March  6.2 (UTC)
with no trace of a host galaxy.  The NUV--MAMA prism spectrum presents
a  relatively flat   spectrum    (in $f_\lambda$) between    2800  and
\mbox{3300  \AA}, with  a mean flux $8.7^{+0.8}_{-1.6}\pm2.6~10^{-18}~
\mathrm{ergs}~\mathrm{s}^{-1}~\mathrm{cm}^{-2}~\mbox{\AA}^{-1}$  and a
sharp  break centered at $2797\pm25$ \AA,  interpreted as a \ion{H}{1}
Lyman break.  It indicates that the OT arises in or beyond an absorber
at $z = 2.067\pm0.025$  similar to  the  ones causing the  high column
density systems  detected in  QSO spectra.  The redshift  and the high
\ion{H}{1} column   density  are      confirmed  by  the      observed
$\log{N_\mathrm{HI}} =  21.2\pm0.5$ damped Ly-$\alpha$,   \ion{Fe}{2},
\ion{C}{4}  and  other    lines   observed in  the    ESO/VLT spectrum
\citep{jfg+00},  as well as the \ion{Mg}{2}   and \ion{Fe}{2} lines in
the Keck  spectrum \citep{cas00}.  The  lack of a detected host galaxy
and   the large \ion{H}{1}  column  density system  are similar to the
situation   for   damped   Ly-$\alpha$     systems  at   $z   >     2$
\citep[e.g.][]{lhg+95,fmw99}, which are thought to be mainly caused by
merging   protogalactic   clumps hosted  by  collapsed   DM  halos and
consequently probe  the  progenitors of  normal  present--day galaxies
\citep{gkh+97,hsr98,hsr00}. Combined with the fact  that other OTs are
often found in star forming galaxies, presumably with large \ion{H}{1}
content, this strongly suggests that  $z = 2.067\pm0.025$ is  actually
the redshift of \objectname[]{GRB~000301C}.  If this  is the case, our
spectrum as well  as the ESO/VLT  and the  Keck ones  suggest that the
line-of-sight to  the OT crosses the interstellar  medium  of its host
galaxy. 

Our data are  compatible  with no or  little extinction.  The absolute
magnitude of the  OT at the time  of  the discovery  image, 41.5 hours
after   the  burst, was therefore  $M_\mathrm{R}   = -26.1$ for $H_0 =
65~\mbox{km}~\mbox{s}^{-1}~     \mbox{Mpc}^{-1}$,  $\Omega =  0.3$ and
$\Lambda =  0.7$.  The  low  extinction would   argue against a  naive
interpretation  that $\gamma$-ray  bursts originate in  a star forming
region.  Instead, one  might  imagine this favors the  hypothesis that
\objectname[]{GRB~000301C} results from a neutron star -- neutron star
or neutron star  -- black hole binary  merger \citep{pac91,nar92}.  In
this case, this relatively short duration event could have taken place
at  a  significant distance from  the  host galaxy some  time  after a
significant burst of  star formation.  \citet{nar92} predict that  the
median distance  between a  binary  merger  and  its  host galaxy   is
$\sim50$ kpc  at  the time  of the   burst.   Consequently, it  is not
impossible  that  the  nearby  galaxy    $\sim 2\arcsec$  to  the   NE
\citep{rho00,kgt+00} is actually the  host,  although it is very   red
where as previous   hosts  tend to be   blue in  optical  -- infra-red
colors: we derive a $R - K'$ color  of $4.9\pm0.5$ based on magnitudes
measured by \citet{rho00} and \citet{vei00}.  In particular, we cannot
exclude  that this galaxy  lies at $z \sim 2.067$  on the basis of the
existing photometry.   If this is  the case,  its \ion{H}{1} disk must
extend  over     at         least $\sim    20    $kpc        ($H_0   =
65~\mbox{km}~\mbox{s}^{-1}~\mbox{Mpc}^{-1}$,    $\Omega_0 =  0.3$  and
$\Lambda = 0.7$) to cover the line-of-sight of  the OT.  This value is
similar  to  the separation  between   the  line-of-sight to  the  QSO
Q~2233+131 and the galaxy  identified by \citet{dpb+96} as responsible
for the $z = 3.150$ damped Ly-$\alpha$ in  its spectrum.  Spectroscopy
of the galaxy $\sim 2\arcsec$  to the NE of \objectname[]{GRB~000301C}
is necessary to settle the issue.

However,    the  present data cannot   exclude   that the burst itself
destroys dust as high-energy  photons find their  way  out of  a dense
cloud \citep{wax00}.  On the other  hand, the time elapsed between the
formation of the GRB progenitor and the GRB event itself could be long
enough for  the progenitor to leave  regions of large extinction as is
the case for core-collapse  supernovae (type IIa, Ib/c), a significant
fraction  of which     are    also little  affected     by  extinction
\citep[e.g.][]{ske+94}.  Finally, the low extinction could be possibly
explained if the progenitor  of \objectname[]{GRB~000301C} was located
in a low metallicity galaxy which has only recently started to produce
stars. The host of \objectname[]{GRB~000301C} might therefore resemble
the galaxies giving rise to damped Ly-$\alpha$ absorbers.

\textit{Note:}  After   this    paper  was    submitted, the    direct
determination of the redshift for the optical counterparts of two other
GRBs have been reported with redshifts similar or  larger than the one
of \objectname[]{GRB~000301C}: \objectname[]{GRB~000926} at $z = 2.066$
\citep{fyn00b} and \objectname[]{GRB~000131} at $z=4.50$
\citep{ahp+00}.

\acknowledgments{
  This work was supported by NASA Guaranteed Time Observer funding to
  the STIS Science Team and is based upon observations obtained with
  the NASA/ESA Hubble Space Telescope, which is operated by the
  Association of Universities for Research in Astronomy, Inc., under
  NASA contract NAS 5-26555.  This work was supported in part by NASA
  through STIS GTO funding. 
by STScI GO funding  under NASA contract NAS 5-26555 and by the
Danish Natural Research Council (SNF). KH is grateful to JPL Contract
958056 for Ulysses support, and to NAG 5 9503 for NEAR support.
}

\clearpage 

\begin{center}
  \leavevmode
  \psfig{file=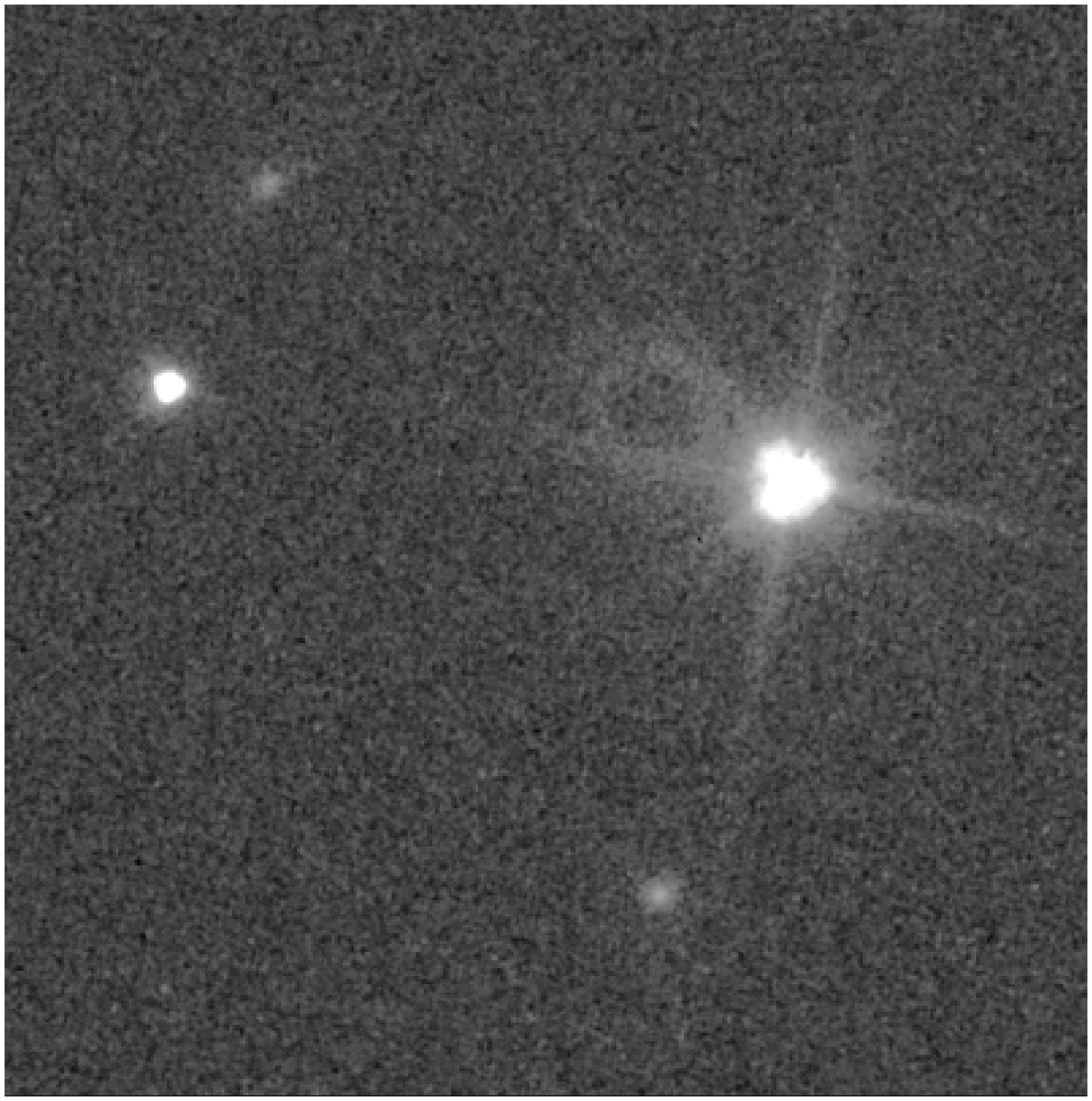,height=3.0in}
\end{center}
\figcaption[f1.eps]{A 10\arcsec$\times$10\arcsec\, portion of the
  \textit{HST}/STIS clear aperture CCD image of the field of
  GRB~000301C showing the OT on the left and the offset
  star used for acquisition on the right.  North is up and East is to
  the left. The ring visible on the left of the offset star is a ghost
  due to internal reflections in the STIS CCD window.
\label{fig:image}
}
\clearpage
\begin{center}
  \leavevmode
  \psfig{file=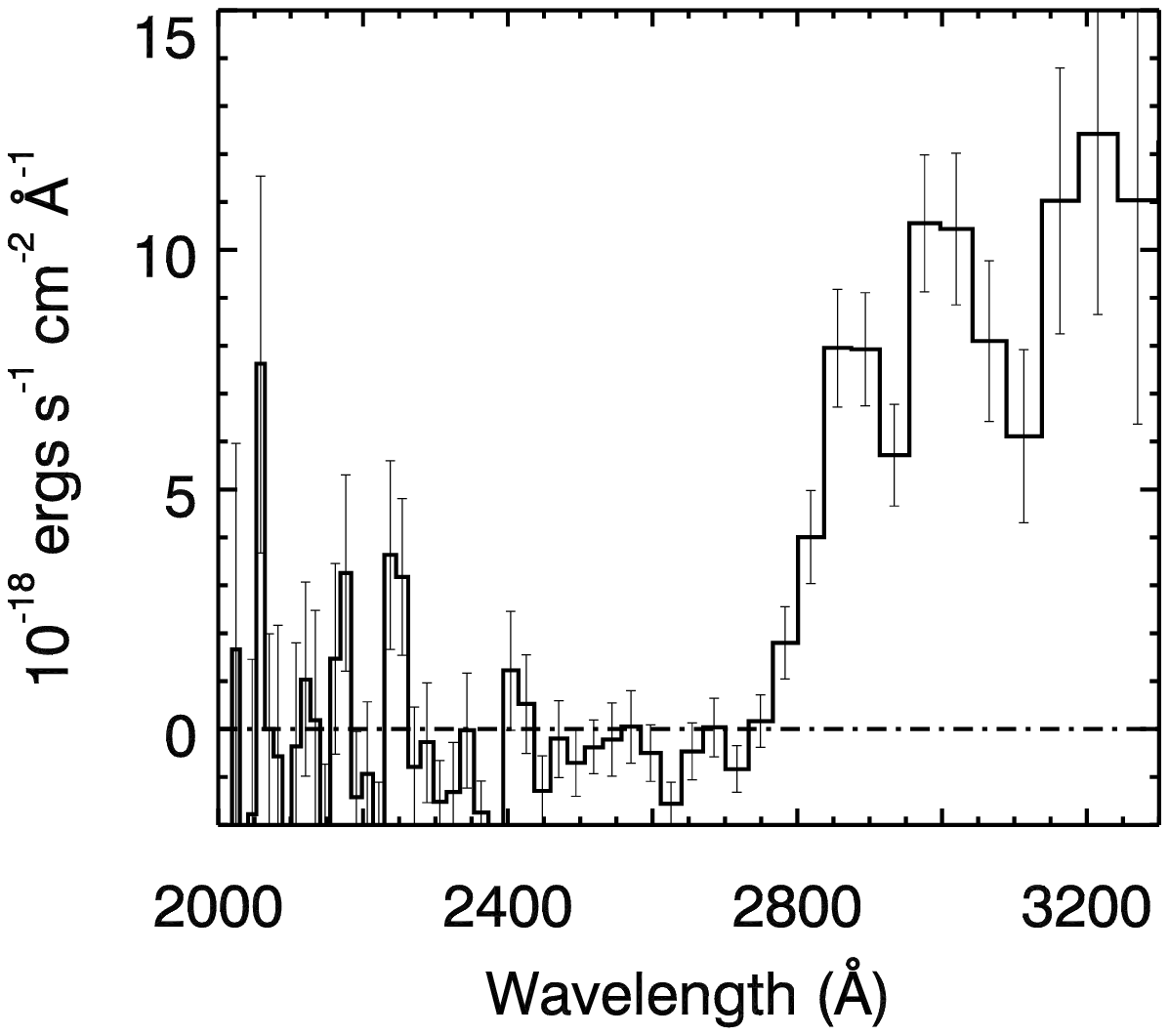,height=2.5in}
\end{center}
\begin{center}
  \leavevmode
  \psfig{file=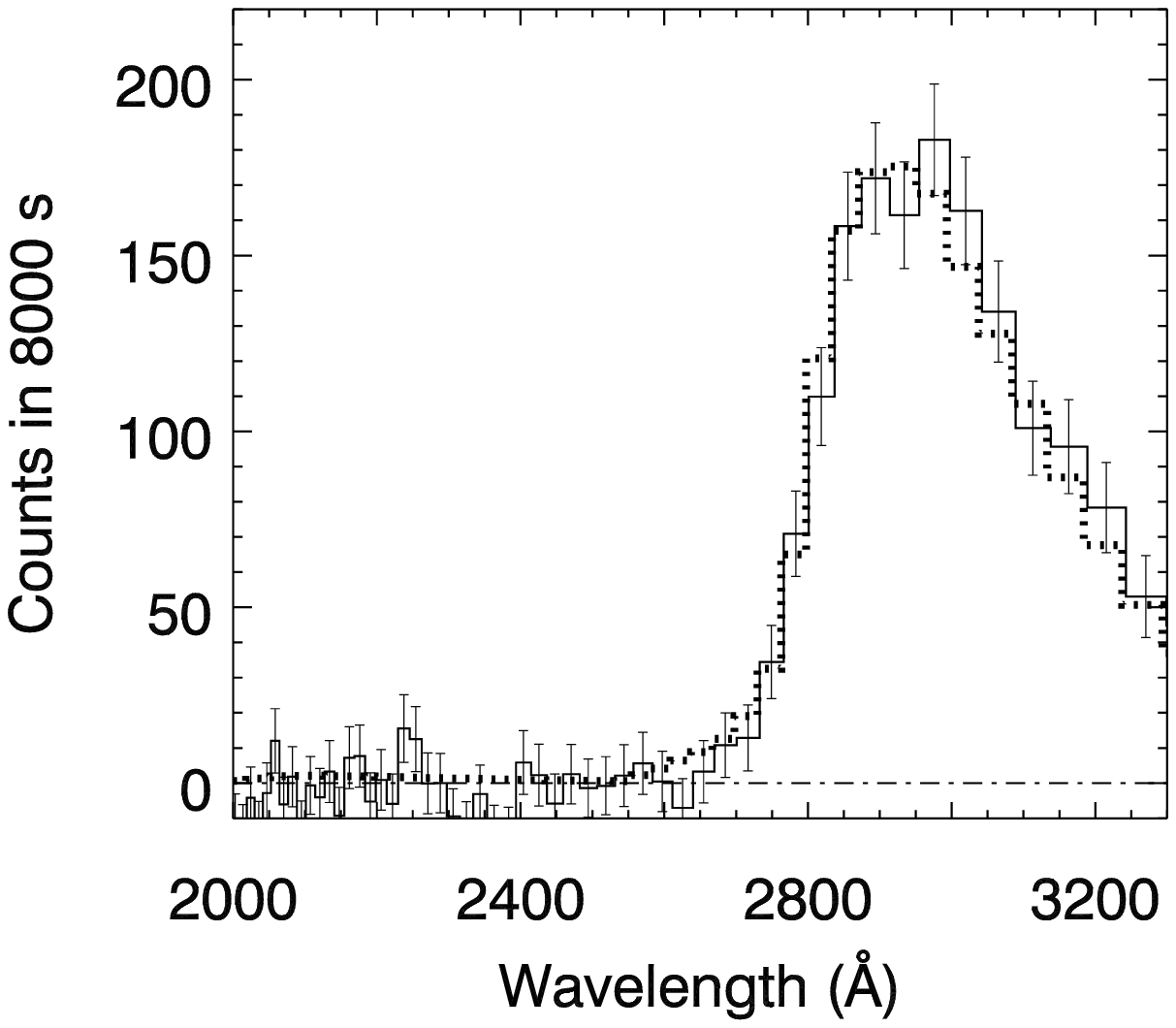,height=2.5in}
\end{center}
\figcaption[f2a.eps,f2b.eps]{(a) \textit{Top:} Deconvolved, flux
  calibrated, UV spectrum of GRB~000301C. A break is clearly seen at
  $\lambda 2797$ \AA. If caused by the onset of Lyman continuous
  absorption due to \ion{H}{1} gas associated with the host galaxy,
  the OT redshift is $z = 2.067\pm0.025$. (b) \textit{Bottom:}
  Comparison of the observed UV spectrum, expressed in observed total
  number of counts with simulated spectra.  The dotted curve
  corresponds to a model with $\alpha = 0.5$ and extinction by a
  SMC--like dust with $A_\mathrm{V} = 0.15$. A model with $\alpha =
  1.2$ and no extinction would present a spectrum indistinguishable in
  this wavelength range from the one shown here.
  \label{fig:simulation}}

\end{document}